# Comments on the papers recently published by Kalaivani et al

Bikshandarkoil R. Srinivasan, Kedar U. Narvekar
Department of Chemistry, Goa University, Goa 403206, India
Email: srini@unigoa.ac.in

**Abstract**

We argue that the so-called *L*-lysinium succinate **1** (D. Kalaivani et al, J. Cryst. Growth 426 (2015) 135-140), zinc chloride doped *L*-lysinium succinate **2** (D. Kalaivani et al, J. Cryst. Growth 428 (2015) 24-28), *L*-threonine phthalate **3** (J. Elberin Mary Theras, D. Kalaivani et al J. Cryst. Growth 427 (2015) 29-35) and sodium acetate doped *L*-tyrosine **4** (D. Arthi et al Spectrochim. Acta, 136A (2015) 168–174) are all dubious crystals. Taking compounds **1**-**4** as examples we show that EDAX is an inappropriate method for characterization of new materials based on elemental composition data.

**Introduction**

Study of amino acids and their compounds is a frontier area of research. A comprehensive and up to date account of many amino acid salts is described in a recent book by Fleck and Petrosyan [1]. In addition to several valuable works by several research groups worldwide, the literature of amino acids unfortunately contains many erroneous reports in the area of crystal growth of amino acid based compounds. Several papers reporting pure amino acids or a well-known compound under the name novel nonlinear optical (NLO) crystal or semiorganic NLO crystal or novel optoelectronic material have been criticized by the research groups of Fleck, Petrosyan, Natarajan, Priolkar, Srinivasan, and Tylczyński [2-15]. The many difficulties associated with the growth and characterization of amino acid based crystalline materials, were first described in an extensive case study [2]. Despite the publication of the case study many erroneous papers continue to appear in the literature, for example the papers by the authors of [16-19] reporting growth of so called *L*-lysinium succinate, zinc chloride doped *L*-lysinium succinate, *L*-threonine phthalate and sodium acetate doped *L*-tyrosine crystals. All these supposedly novel compounds are referred to by names and abbreviated by unusual codes like LLMS **1**, (ZnCl$_2$-Lls) **2** and so on. In addition to showing that compounds **1**-**4** are dubious crystals and not novel nonlinear optical (NLO) crystals as claimed in [16-19], we show that EDAX is an inappropriate technique for quantitative determination of elemental composition.

**Comment**

The authors claim to have obtained crystals of *L*-lysinium succinate **1** [16], zinc chloride doped *L*-lysinium succinate, **2** [17] and *L*-threonine phthalate **3** [18] by the slow evaporation of a solution containing *L*-lysine and succinic acid (for **1**) in water, *L*-lysine hydrochloride, succinic acid and ZnCl$_2$ (for **2**) in water and *L*-threonine and phthalic acid (for **3**) in methanol-acetic acid mixed solvent and L-tyrosine and sodium acetate (for **4**) in mixed solvent of HCl and water. The crystal growth procedures which do not contain details of the synthesis in terms of quantities of reagents taken, volume of solvent used and yield of product obtained, are not acceptable. Although authors claim to have performed single crystal X-ray experiments, the formulae C$_{10}$H$_{20}$N$_2$O$_6$, C$_{10}$H$_{20}$N$_2$O$_6$·ZnCl$_2$ and C$_{11}$H$_{14}$NO$_5$Na were reported for compounds **1**, **2** and **4** respectively based on a chemical equation for the crystal growth reaction while no formula is given in the entire paper for **3**. The chemical equation reported for the formation of compound **2** showing evolution of HCl gas from *L*-lysine hydrochloride on reaction with aqueous ZnCl$_2$ appears not only questionable but also arbitrary without any chemistry reasoning. The dubious nature of all crystals can be evidenced from their names. For example, the name *L*-lysinium succinate is unacceptable because *L*-lysinium is a monocation and succinate is a dianion, indicating that **1** is a charge imbalanced crystal. It is not clear if the authors are actually referring to the 1:1 *L*-lysinium hydrogen succinate or the 2:1 bis(*L*-lysinium) succinate. Similarly the name *L*-threonine phthalate **3** appears to be that of a charge imbalanced solid because *L*-threonine is a neutral molecule while phthalate is a dianion. It is not clear why **2** and **4** are referred to as Zn doped or Na doped crystals respectively despite the proposed formula containing one Zn (or Na) per formula unit.

The assignment of the centrosymmetric monoclinic space group *C2/c* for a solid supposed to contain a chiral amino acid confirms beyond doubt that **3** is a dubious crystal. The same argument can be extended for compound **2** also supposed to contain *L*-lysinium component, for which the authors reported the centrosymmetric monoclinic space group *P2$_1$/c*. For all four compounds the authors have not reported any CIF data to substantiate the findings of their single crystal work but instead only unit cell parameters indicating the questionable nature of their findings. It is quite unfortunate to note that authors reported a snapshot of computer screen in the form of a figure with the caption '*XRD pattern of LLMS* for **1** and *single crystal XRD pattern with cell parameters* for **3**. It is regrettable to note such figures can be published in a peer reviewed paper and this raises serious concern about this series of papers by the group of Joseph [16-19]. The non-adherence to standard procedures of reporting single crystal data, assignment of centrosymmetric space groups for compounds supposedly containing chiral amino acid fragments and reporting snap shots of computer screen only reveal that the authors do not have the necessary expertise to interpret and report single crystal data. Although the authors believe that their so called *L*-lysinium succinate is a novel non-linear optical material they were unaware that the *L*-lysine-succinic acid system has been well studied by Prasad and Vijayan [20]. The authors also did not consider it important to compare their unit cell parameters with the reported data for the starting materials or related compounds used for crystal growth. Had they done this exercise they would have readily noted that the cell parameters of **3** are in excellent agreement with those of phthalic acid [21] for a so called LTP and with those of *L*-tyrosine hydrochloride for **4** [22].

In this work we have calculated the theoretical % for each element based on the authors proposed formula (Table 1) for comparison with experimental EDAX data. Our calculation is correct as evidenced by the fact that all % add to 100. Despite literature reports showing that EDAX cannot be used to get quantitative elemental analytical data [10-12], this technique is more frequently and incorrectly used [23] for new compound characterization as observed in all the commented papers [16-19]. In all these papers the authors claimed "*The chemical composition is confirmed quantitatively by EDAX analysis*". While making the claim the authors did not calculate the expected percentages for the proposed formula as they did not take into consideration that the composition obtained from EDAX study does not contain any H content. Unfortunately, the authors are unaware that EDAX is an inappropriate method to determine the quantitative composition of lighter elements like H, Li etc. Actually, the reported EDAX data for all compounds in Table 1 convincingly prove that all are dubious crystals. A scrutiny of the reported data reveals that the total % of all compounds (without any H) amount to 100% and the other reported data do not in any way correspond to the expected (theoretically calculated) values for the proposed formula. The absurdity of the claims can be evidenced for the amino acid based compound **2** not having any N. The so called doped Zn (or Na) compounds contain much less % Zn (or Na) than expected for the proposed formula. In the case of compound **3** where authors did not report any formula, the experimental EDAX data are neither in agreement with the expected values for a 1:1 or a 2:1 compound of *L*-threonine and phthalic acid. However the presence of N for this compound which should be actually pure phthalic acid (based on the unit cell) only indicates that **3** is impure phthalic acid. In view of the dubious nature of **1** to **4** all other studies on such crystals are meaningless and do not deserve any mention.

**Table 1.** Calculated elemental analytical data for compounds **1** to **5** based on molecular formula proposed by authors

| Compound | Formula weight | %C | %H | %N | %O | %X* | %M# | Total % |
|---|---|---|---|---|---|---|---|---|
| So called LLMS **1** $C_{10}H_{20}N_2O_6$ | 264.28 | 45.45 (**37.07**) | 7.63 --- | 10.60 (**27.13**) | 36.32 (**35.82**) | --- --- | --- --- | 100.0 100.02 |
| So called LTP **3** $C_{12}H_{15}NO_7$ | 285.25 | 50.53 (**45.97**) | 5.30 --- | 4.91 (**19.75**) | 39.26 (**34.28**) | --- --- | --- --- | 100.0 100.0 |
| bis(*L*-threoninium) phthalate | 404.37 | 47.52 | 5.98 | 6.93 | 39.57 | --- | --- | 100.0 |
| So called ZnCl$_2$-Lls **2** $C_{10}H_{20}N_2O_6 \cdot ZnCl_2$ | 400.57 | 29.98 (**54.26**) | 5.04 --- | 6.99 --- | 23.97 (**42.57**) | 17.70 (**1.43**) | 16.32 (**1.74**) | 100.0 100.0 |
| So called LTSA **4** $C_{11}H_{14}NO_5Na$ | 263.22 | 50.19 (**63.09**) | 5.37 --- | 5.32 (**8.87**) | 30.39 (**27.86**) | --- --- | 8.73 (**0.19**) | 100.0 100.0 |
| *L*-proline lithium bromide monohydrate $C_5H_{11}LiNO_3Br$ **5** | 219.99 | 27.30 (**30.38**) | 5.04 --- | 6.37 (**7.57**) | 21.82 (**27.22**) | 36.31 (**34.83**) | 3.16 --- | 100.0 100.0 |

* X = Cl for **2** and Br for **4**; M = Zn for **2,** Na for **4** and Li for **5**; **Values in bracket** are experimental data from EDAX reported for **1-5** by authors of [16-19 & 23]; For LTP, formula $C_{12}H_{15}NO_7$ corresponds to the 1:1 product namely *L*-threoninium hydrogenphthalate.

In this context mention is made of a so called *L*-proline lithium bromide monohydrate **5** reported by Shkir et al [23]. For **5** the authors performed a single crystal study and determined its structure and also reported EDAX data without any H and Li content with the total % of C, N, O and Br adding to 100%. Despite this and the data for C, N and Br not in agreement with expected values, the authors claimed "*The EDXS analysis confirms the synthesis of new compound*" [23]. The above only means that the authors did neither calculate the theoretical % nor are they aware of the limitations of the EDAX technique. The best explanation for the EDAX data is that the compound under study contains C, N and Br but no Li or H.

Although the exact purpose of the EDAX study of **5** is not very clear, the conflicting single crystal and EDAX data only indicates **5** is a dubious crystal and not *L*-proline lithium bromide monohydrate 5. In all the cases, the reason for the mismatch of the experimental data with the theoretical values is due to the inability of EDAX to detect lighter elements like H or Li and considering the % of the other elements viz. equal to 100%. In view of this the EDAX technique should not be used as a confirmatory proof for product characterization of new compounds containing lighter elements like H, Li etc.

**Conclusions**

New compounds should not be referred to by arbitrary names abbreviated by strange codes. The observation of the presence of a few elements in an EDAX study is not a confirmatory proof for the formula of a new compound. The compounds **1-5** reported by the authors of the commented papers are dubious crystals.

Footnote: Compounds are referred to by numbers to avoid use of non-standard codes

**References**

[1] M. Fleck, A.M. Petrosyan, Salts of amino acids: crystallization, structure and properties. Springer, Dordrecht, 2014.

[2] M. Fleck, A.M. Petrosyan, Difficulties in the growth and characterization of non-linear optical materials: a case study of salts of amino acids. J. Crystal Growth, 312, (2010) 2284-2290. http://dx.doi.org/10.1016/j.jcrysgro.2010.04.054

[3] A.M.Petrosyan, Comment on ''Growth and characterization of glycine picrate single crystal'' by T. Uma Devi et al. [Spectrochim. Acta A71 (2008) 340–343], Spectrochimica Acta A75 (2010) 1176. http://dx.doi.org/10.1016/j.saa.2009.12.045

[4] V.V. Ghazaryan, M. Fleck, A.M. Petrosyan, Glycine glycinium picrate—Reinvestigation of the structure and vibrational spectra, Spectrochimica Acta A78 (2011) 128–132. http://dx.doi.org/10.1016/j.saa.2010.09.009

[5] A.M. Petrosyan, V.V. Ghazaryan, M. Fleck, On the existence of "bis-glycine maleate" J. Crystal Growth 359 (2012) 129-131. http://dx.doi.org/10.1016/j.jcrysgro.2012.08.024

[6] A.M. Petrosyan, V.V. Ghazaryan, M. Fleck, On the existence of "*L*-threonine formate", "*L*-alanine lithium chloride" and "bis *L*-alanine lithium chloride" crystals, Spectrochimica Acta A105, 623-625 (2013). http://dx.doi.org/10.1016/j.saa.2013.01.009

[7] Z. Tylczyński, M. Wiesner, Comment on "Ferroelectricity in glycine picrate: An astonishing observation in a centrosymmetric crystal" [Applied Physics Letters 95, 252902 (2009)], Applied Physics Letters 96 (2010) 126101. http://dx.doi.org/10.1063/1.3372622

[8] Z. Tylczyński, Comment on "Effect of rare earth ions on the properties of glycine phosphate single crystals" by K. Senthilkumar et al. [J. Cryst. Growth 362 (2013) 343–348], J. Crystal Growth, 382 (2013) 94-95. http://dx.doi.org/10.1016/j.jcrysgro.2013.08.002

[9] B.R. Srinivasan, K.R. Priolkar, Comment on 'Synthesis, growth, structural, spectroscopic, crystalline perfection, second harmonic generation (SHG) and thermal studies of 2-aminopyridinium picrate (2APP): A new nonlinear optical material' [Solid State Sci. 14 (2012) 773-776], Solid State Sciences 20 (2013) 15-16. http://dx.doi.org/10.1016/j.solidstatesciences.2013.03.005

[10] B.R. Srinivasan, P. Raghavaiah, V.S. Nadkarni, Reinvestigation of growth of urea thiosemicarbazone monohydrate crystal, Spectrochimica Acta, A112 (2013), 84–89. http://dx.doi.org/10.1016/j.saa.2013.04.026

[11] B.R. Srinivasan, T. A. Naik, Z. Tylczyński, K.R. Priolkar, Reinvestigation of growth of thiourea urea zinc sulfate crystal, Spectrochimica Acta, 117A (2014), 805–809. http://dx.doi.org/10.1016/j.saa.2013.08.083

[12] B.R. Srinivasan, S. Natarajan, G.N. Gururaja, Comments on the papers recently published by Sangeetha et al. J. Thermal Analysis & Calorimetry 120 (2015) 1071-1075.

[13] A.M. Petrosyan, B.R. Srinivasan, On the existence of "*L*-arginine acetamide", J. Thermal Analysis & Calorimetry 120 (2015) 1077-1078. http://dx.doi.org/10.1007/s10973-015-4530-3

[14] B.R. Srinivasan, K. Moovendaran, S. Natarajan, Comments on ''Crystal growth, spectral, optical, and thermal characterization of glycyl-L-alanine hydrochloride (GLAH) single crystal, J. Thermal Analysis & Calorimetry, 118 (2014) 1397-1399. http://dx.doi.org/10.1007/s10973-014-4164-x

[15] A.M. Petrosyan, V.V. Ghazaryan, Z. Tylczyński, B.R. Srinivasan, On the existence of "a new ferroelectric tri-glycine barium nitrate", Ferroeclectrics Letters Section (2015) (**In press**) http://dx.doi.org/10.1080/07315171.2015.1068970

[16] D. Kalaivani, D. Arthi, A. Mukunthan, D. Jayaraman, V. Joseph, Growth and characterization of *L*-lysinium succinate single crystal. J. Crystal Growth 426, (2015) 135-140. http://dx.doi.org/10.1016/j.jcrysgro.2015.05.030

[17] D. Kalaivani, D. Jayaraman, V. Joseph, Investigations on 2D and 3D topography and Z-scan studies of zinc chloride co-doped L-lysinium succinate. J. Crystal Growth 428 (2015) 24-28. http://dx.doi.org/10.1016/j.jcrysgro.2015.07.017

[18] J. Elberin Mary Theras, D. Kalaivani, D. Jayaraman, V. Joseph, Growth and spectroscopic, thermodynamic and nonlinear optical studies of *L*-threonine phthalate crystal. J. Crystal Growth 427 (2015) 29-35. http://dx.doi.org/10.1016/j.jcrysgro.2015.06.009

[19] D. Arthi, D. Anbuselvi, D. Jayaraman, J. Arul Martin Mani, V. Joseph, Nucleation, growth and characterization of semiorganic nonlinear optical crystal sodium acetate doped l-tyrosine" Spectrochimica Acta, 136A (2015),168–174. http://dx.doi.org/10.1016/j.saa.2014.08.117

[20] G. Sridhar Prasad, M. Vijayan, X-ray studies on crystalline complexes involving amino acids and peptides. XXIII. Viriability in ionization state, conformation and molecular aggregation in the complexes of succinic acid with *DL*- and *L*-lysine. Acta Crystallogr. B47, 927-935 (1991). http://dx.doi.org/10.1107/S0108768191004962

[21] O. Ermer, Ungewöhnliches Strukturmerkmal kristalliner Phthalsäure. Helvetica Chimica Acta 64, 1902-1909 (1981). http://dx.doi.org/10.1002/hlca.19810640623

[22] M.N. Frey, T.F. Koetzle, M.S. Lehmann and W.C. Hamilton, Precision neutron diffraction structure determination of protein and nucleic acid components. X. A comparison between the crystal and molecular structures of *L*-tyrosine and *L*-tyrosine hydrochloride J. Chemical Physics 58, (1973) 2547 http://dx.doi.org/10.1063/1.1679537

[23] M. Shkir, S. Alfaify, M.A. Khan, E. Dieguez, J. Perles, Synthesis, growth, crystal structure, EDX, UV–vis–NIR and DSC studies of *L*-proline lithium bromide monohydrate—A new semiorganic compound, J. Crystal Growth, 391 (2014) 104-110. http://dx.doi.org/10.1016/j.jcrysgro.2014.01.012